\documentclass[12pt, titlepage, reqno]{article}
\usepackage{authblk}
\usepackage[margin=1in]{geometry}
\RequirePackage{amsthm,amsmath,amsfonts,amssymb}
\usepackage{array}
\RequirePackage{natbib}
\RequirePackage[colorlinks,citecolor=blue,urlcolor=blue]{hyperref}
\usepackage{booktabs,longtable}
\usepackage{tabularx}
\usepackage{multirow}
\usepackage{threeparttable}
\usepackage{graphicx}
\graphicspath{{./}{../image/}}
\usepackage{xcolor}
\usepackage{soul}
\usepackage{epstopdf}
\usepackage[inline]{enumitem}
\usepackage{bm}
\usepackage[ruled, lined]{algorithm2e}
\usepackage{setspace}
\usepackage{float}
\usepackage{rotating}
\usepackage{subcaption}

\definecolor{steelblue}{RGB}{70, 130, 180}

\title{When Does the Silhouette Score Work? A Comprehensive Study in 
Network Clustering}

\author[1]{Zongyue Teng}

\author[2]{Jun Yan}

\author[1]{Dandan Liu}

\author[1,*]{Panpan Zhang}

\affil[1]{Department of Biostatistics, Vanderbilt University Medical 
Center, Nashville, TN 37203, USA}

\affil[2]{Department of Statistics, University of Connecticut, 
Storrs, CT 06269, USA}

\affil[*]{Correspondence: \href{mailto:panpan.zhang@vumc.org}{Panpan 
Zhang}}

\begin{document}
	
\maketitle
	
\begin{abstract}
Selecting the number of communities is a fundamental challenge in 
network clustering. The silhouette score offers an intuitive, 
model-free criterion that balances within-cluster cohesion and 
between-cluster separation. Albeit its widespread use in clustering 
analysis, its performance in network-based community detection 
remains insufficiently characterized. In this study, we 
comprehensively evaluate the performance of the silhouette score 
across unweighted, weighted, and fully connected networks, examining 
how network size, separation strength, and community size imbalance 
influence its performance. Simulation studies show that the 
silhouette score accurately identifies the true number of 
communities when clusters are well separated and balanced, but it 
tends to underestimate under strong imbalance or weak separation and 
to overestimate in sparse networks. Extending the evaluation to a 
real airline reachability network, we demonstrate that the 
silhouette-based clustering can recover geographically interpretable 
and market-oriented clusters. These findings provide 
empirical guidance for applying the 
silhouette score in network clustering and clarify the conditions 
under which its use is most reliable.

\bigskip
\noindent{\bf Keywords.}
Community detection;
Simulation study;
Stochastic block model;
Weighted networks
\end{abstract}

\doublespacing

\section{Introduction}

Network clustering has emerged as a powerful analytical approach for 
uncovering modular structures in complex systems. By grouping nodes
according to similarity or connectivity patterns, clustering enables
the identification of meaningful communities in diverse domains,
including biology \citep{pavlopoulos2011using}, neuroscience
\citep{sporns2016modular}, and social sciences
\citep{ouyang2020clique,ouyang2023amixed}. Such communities may
represent gene co-expression modules, brain connectivity networks,
or social subgroups. A central challenge in network clustering and 
community detection is determining the correct number of clusters (or 
communities). In real-world applications, this quantity typically 
lacks a ground truth, but it significantly influences downstream 
inference and interpretation. To date, many widely used network 
clustering approaches (e.g., stochastic block models) require the 
number of clusters to be specified a priori, rendering the accurate 
determination of cluster number a critical first step in network 
analysis.

A variety of methods have been proposed for determining the number of 
clusters in network clustering problems. Heuristic and statistical 
indices include the gap statistic \citep{tibshirani2001estimating}, 
the silhouette score \citep{rousseeuw1987silhouettes}, modularity 
maximization \citep{brandes2008onmodularity}, and eigengap heuristics 
for spectral clustering \citep{vonluxburg2007atutorial}. Model-based 
approaches rely on information criteria such as AIC and BIC
\citep{fraley2002model, spiegelhalter2002bayesian, 
hu2003acomparative}, especially in the context of stochastic block 
models~\citep{abbe2018community} and their
extensions. Numerous alternative validity measures have also been
proposed \citep{dunn1973fuzzy, calinski1974dendrite, 
davies1979cluster, xie1991validity}. Despite these advances, to
date, no single method has consistently outperformed others across
diverse applications \citep{arbelaitz2013extensive,
  liu2010understanding}.

The silhouette score is particularly attractive because it quantifies 
how well an observation fits within its assigned cluster relative to 
other clusters, without relying on explicit parametric models or 
reference 
distributions~\citep{rousseeuw1987silhouettes,shahapure2020cluster}. 
It has been widely used in applications such as neuroimaging
\citep{mwangi2014visualization,ryali2015development,
grossberger2018unsupervised} and single-cell RNA sequencing
\citep{xu2015identification,kiselev2017sc3}. For a given clustering 
assignment, the silhouette score evaluates its performance by jointly 
measuring within-cluster cohesion and between-cluster separation, 
enabling direct comparison across different numbers of clusters to 
guide selection of the cluster number~\citep{pavlopoulos2011using}.

Empirical studies, largely based on simulation experiments and 
benchmark datasets with known cluster structure, have shown that the 
average silhouette score across observations can provide a promising 
criterion for selecting the number of clusters across a range of 
dimensionalities \citep{starczewski2015performance}.
Nonetheless, studies have also shown that its performance
deteriorates when clusters overlap or are weakly separated,
leading the silhouette index to underestimate the true 
number~\citep{arbelaitz2013extensive}. Comparative evaluations of 
internal cluster validation indices (which assess clustering 
structure using only the observed data) indicate that the silhouette 
score generally performs favorably relative to many alternative 
criteria~\citep{arbelaitz2013extensive}, such as the Dunn index, the 
Calinski--Harabasz index, and the Davies--Bouldin index, though its 
reliability diminishes in the presence of noise or closely
connected subclusters \citep{liu2010understanding}. Most existing 
comparative evaluations focus on traditional clustering settings, 
such as $k$-means or hierarchical clustering applied to vector-valued 
data. Comprehensive investigations of the performance of silhouette 
score in network-based clustering, where similarity is
captured by graph structures rather than direct feature
representations, remain limited. Given the growing use of network 
clustering in complex systems, including biological pathways, social 
networks, and brain architecture, this gap motivates further 
investigation into the strengths and limitations of the silhouette 
score for selecting the number of communities in network settings.

This study aims to fill this research gap and makes two major 
contributions. First, we evaluate the performance of the silhouette 
score in unweighted networks and weighted networks, respectively 
modeled by stochastic block models and their weighted extensions. 
Second, we assess its robustness to network size, community 
separation, and imbalance in community sizes. In addition, we 
complement the simulation studies with a real-data case study, 
demonstrating the application of the silhouette score to an airline 
reachability network. Together, the findings of this study provide 
practical guidance for applying silhouette scores in community 
detection and highlight open challenges for future methodological 
development in network clustering analysis.

The remainder of the manuscript is organized as follows.
Section~\ref{sec:silhouette} introduces the silhouette score in the 
context of community detection for network data.
Section~\ref{sec:sim} presents simulation studies evaluating its 
performance under a variety of network configurations.
Section~\ref{sec:air} provides a real-data case study 
illustrating silhouette-based module detection in an airline 
reachability network. Finally, Section~\ref{sec:discuss} concludes 
with a summary of key findings and discusses potential directions 
for future research.

\section{Silhouette Score}
\label{sec:silhouette}

The silhouette score provides a model-free index of clustering
quality by balancing cohesion within clusters against separation
between clusters. In network settings, the measure is applied to
distances derived from the graph structure rather than feature
vectors. This section introduces its formal definition in the
network context and then outlines major limitations that motivate
the comprehensive evaluation in later sections.

\subsection{Definition}

A network and its adjacency matrix define the structure on which
clustering is based. Let $G := (V, E)$ denote an undirected,
weighted network with node set $V$ (where $|V| = n$ denotes the 
network size), edge set $E$, and adjacency matrix $\bm{W} := 
(w_{ij})_{n \times n}$, where $w_{ij}$ is the weight of the 
edge between nodes $i$ and $j$, with $w_{ij}$ = 0 indicating no 
edge. Self-loops are excluded, so $w_{ii} = 0$ for all $i$. The 
network structure induces a distance matrix $\bm{D} = (d_{ij})_{n 
\times n}$, where $d_{ij}$ represents a user-defined measure of 
dissimilarity between nodes~$i$ and $j$. By convention, $\bm{D}$ is 
constructed from the adjacency matrix$\bm{W}$. In our simulation 
study, because the edge weights lie in the interval $[0,1]$, we 
define the dissimilarity as $d_{ij} = 1 - w_{ij}$. This choice is not 
unique; more generally, any well-defined distance metric derived from 
the network topology and, when available, edge weights may be used, 
depending on the application and desired interpretation. Suppose $K$ 
is the number
of clusters in $G$, and let
$\bm{Z}_i = (Z_{i1}, \ldots, Z_{iK}) \in \{0,1\}^K$
denote the membership of node $i$, subject to
$\sum_{k=1}^K Z_{ik} = 1$, indicating unique membership for each 
node. A clustering result is then represented
by $\bm{Z} = \{\bm{Z}_i : i = 1, \ldots, n\}$, the collection of
memberships across all nodes.

Given a clustering result, the silhouette score evaluates cohesion 
and separation through within- and between-cluster distances. The
within-cluster distance of node $i$ is
\begin{equation}
  \label{eq:sil_a}
  a_i = \frac{1}{n_k - 1} \sum_{j \ne i: Z_{jk} = 1} d_{ij},
\end{equation}
where $n_k = \sum_{j=1}^n Z_{jk}$ is the size of cluster $k$. Smaller 
values of $a_i$ indicate greater 
cohesion. The between-cluster distance of node~$i$ is
\begin{equation}
  \label{eq:sil_b}
  b_i = \min_{\ell \neq k}\left(\frac{1}{n_\ell}
  \sum_{j: Z_{j\ell} = 1} d_{ij}\right),
\end{equation}
which reflects the minimum average dissimilarity from node $i$ to
another cluster. Larger values of $b_i$ represent stronger
separation. Together, $a_i$ and $b_i$ provide the foundation for the
silhouette score.

The silhouette score combines cohesion and separation into a
normalized measure of clustering quality. For node~$i$, the score is
\[
  s_i = \frac{b_i - a_i}{\max\{a_i, b_i\}},
\]
which lies between~$-1$ and~$1$. Values near~$1$ indicate a strong
fit within the assigned cluster, values near~$0$ correspond to nodes
on cluster boundaries, and values near~$-1$ suggest possible
misclassification. For singleton clusters with $n_k=1$, $a_i$ is
undefined and~$s_i$ is set to~$0$ by convention. The global score
summarizes the clustering result by averaging node-level scores
across the network,
\[
  s_G = n^{-1} \sum_{i \in V} s_i.
\]
In practice, $s_G$ is often used to compare candidate
clustering results across different values of~$K$, selecting the one 
that maximizes~$s_G$. Although widely applied and extended in recent
work~\citep{lenssen2024medoid, vardakas2024deep}, the performance of 
silhouette score depends on the underlying network structure, and 
certain network configurations may therefore lead to misleading 
results.

\subsection{Strengths and Limitations of Silhouette Score}
\label{sec:limit} 

The silhouette score is a widely used criterion for selecting the 
number of clusters and offers several notable advantages. Because it 
does not rely on parametric model assumptions or pre-specified 
reference distributions, it can be applied broadly across diverse 
data types and clustering methods without modification 
\citep{shahapure2020cluster}. As a normalized measure taking values 
between $-1$ and $1$, the silhouette score is intuitive and easy to 
interpret. Empirical studies based on simulations and benchmark 
datasets further suggest that it can effectively capture clustering 
structure across a range of settings, performing well under varying 
data dimensionality and cluster density 
\citep{arbelaitz2013extensive, liu2010understanding}. In many cases, 
it has been shown to outperform commonly used alternatives, such as 
the Calinski--Harabasz and Davies--Bouldin indices, particularly in 
reflecting differences in cluster separation 
\citep{arbelaitz2013extensive, chicco2025silhouette}. These strengths 
help explain its widespread adoption in practice.

At the same time, the silhouette score depends on distance-based 
cohesion and separation, which can be influenced data 
geometry and network structure. As a result, it may exhibit biases in 
certain settings, potentially leading to inaccurate selection of the 
number of clusters. In the following, we discuss several major 
limitations of the silhouette score that are particularly relevant in 
network clustering and that 
motivate the comprehensive simulation studies presented in this work.

The silhouette score underestimates cluster numbers when clusters are
weakly or moderately separated. In such cases, the score tends to
favor a small number of large clusters, underestimating the true
structure. The mechanism is straightforward: increasing $K$ reduces
the within-cluster distance $a_i$, but also decreases the
between-cluster distance $b_i$ as nearby clusters are introduced.
This compression of the contrast between $a_i$ and $b_i$ lowers the
difference $b_i - a_i$, sometimes producing negative values that
drive down silhouette scores. Larger clusters are less affected,
since they preserve relatively high $b_i$ values even when small
clusters exist nearby. As a result, the score inflates the apparent
quality of merging clusters into fewer groups. In practice, this
means that networks with diffuse community boundaries or overlapping
membership are likely to yield underestimated values of $K$ when
evaluated by the silhouette score, despite the presence of more
complex structure.

The silhouette score performs poorly for non-convex or irregular
cluster shapes. Because it evaluates clustering using pairwise
dissimilarities, the score is most effective when clusters are
convex and well-separated in the chosen metric space,
where small within-cluster distances and large
between-cluster distances align naturally. For non-convex or
irregular shapes, however, this assumption fails. Nodes within the
same cluster may be far apart, inflating~$a_i$, while nodes from
different clusters may be close in Euclidean space, reducing~$b_i$.
As a result, valid non-convex clusters receive underestimated
silhouette scores. Classic examples include nested rings, where inner
rings may be incorrectly merged, and two-moon structures, where
scores improve if each moon is split into convex arcs, leading to
overestimation of~$K$. These outcomes reflect a bias toward
convex-like partitions, causing the silhouette score to obscure
genuine but irregular communities in network data.

The silhouette score is biased toward large clusters when sizes are
imbalanced. Small clusters located near much larger ones are
especially vulnerable, since proximity to the large cluster
depresses~$b_i$ while limited membership keeps~$a_i$ relatively
high. The resulting silhouette values can be small or negative,
mis-characterizing the small cluster as poorly formed despite its
validity. Because the silhouette score averages across nodes rather
than clusters, large clusters contribute more heavily to the global
value, intensifying the bias.
In practice, merging small clusters into nearby large ones
can increase the overall score, creating an artificial preference for
fewer clusters than truly exist. This behavior is particularly
problematic in networks where community sizes vary widely, as large
dominant clusters mask the structure of smaller ones. The net effect
is systematic underestimation of~$K$, reducing the usefulness of the
silhouette score for detecting heterogeneous community configurations
in network-based applications.

\section{Simulation Study}
\label{sec:sim}

In this section, we present a series of simulation studies to 
evaluate the effectiveness of the silhouette score for selecting the 
number of clusters in network data. The simulation settings are 
designed to reflect a broad range of practical scenarios, leading to 
a comprehensive assessment of both the robustness and the limitations 
of silhouette-based cluster number selection under varying network 
structures and clustering characteristics. By examining performance 
across controlled and interpretable regimes, these simulations 
provide new insights into when and why the silhouette score succeeds 
or fails in network clustering. All simulation scripts, including the 
generated random numbers to ensure reproducibility, are available in 
the online supplement.

\subsection{Network Data Generation}
\label{sec:data_gen}

\subsubsection{Unweighted Networks}
\label{sec:unweighted}

Network data are generated from stochastic block 
models~\citep[SBM,][]{holland1983stochastic,abbe2018community} with 
undirected edges. Simulation results are reported separately for 
weighted and unweighted networks. For the unweighted case, we 
evaluate the performance of silhouette-based $K$ selection across 
network sizes $n \in \{240,600\}$ and true cluster numbers 
$K_{\text{true}} \in \{3,8\}$. The design varies within- and 
between-cluster link probabilities, respectively denoted 
$p_{\text{win}}$ and $p_{\text{btw}}$, which jointly govern both 
the signal strength of community structure and the overall level of 
network sparsity; see Table~\ref{tab:sbm_unweighted_prob} for 
details.

Equal- and unequal-sized cluster configurations are also examined. 
For unequal clusters with $K_{\text{true}} = 3$, one cluster contains 
$80\%$ of the nodes while the remaining $20\%$ are evenly divided 
between the other two clusters. For $K_{\text{true}} = 8$, one 
cluster contains $65\%$ of the nodes, with each of the remaining 
clusters containing $5\%$.

To further investigate how localized weakness in separation affects 
clustering, we include an additional condition for which only a 
single pair of clusters is weakly separated (while all other pairs 
remain strongly separated). This setting allows us to isolate the 
effect of one ambiguous cluster boundary on $K$ selection and 
clustering accuracy. Both equal- and unequal-sized cluster 
configurations are considered under this design, with $p_\text{win} = 
0.3$ and $0.6$ examined separately. For all scenarios, including both 
the general designs described above and this additional condition, 
each case is replicated with $200$ simulation runs.

\begin{table}[tbp]
\centering
\renewcommand{\arraystretch}{1.2}
\caption{Within- and between-cluster link probabilities for 
	generating SBM synthetic network data.}
\label{tab:sbm_unweighted_prob}
\setlength{\tabcolsep}{12pt}
\begin{tabular}{c ccccccccc}
	\toprule
	\textbf{$p_{\text{win}} \; \backslash \; 
		p_{\text{btw}}$} & $0.05$ & $0.1$ & $0.15$ & $0.2$ & $0.25$ & 
		$0.3$ & $0.35$ & $0.4$ & $0.45$ \\
	\midrule
	$0.3$ & \checkmark & \checkmark & \checkmark & & & & & & \\
	$0.4$ & & \checkmark & \checkmark  & \checkmark & \checkmark & & 
	& & \\
	$0.5$ & & \checkmark & \checkmark & \checkmark & \checkmark & 
	\checkmark & \checkmark & & \\
	$0.6$ & & \checkmark & \checkmark & \checkmark & \checkmark & 
	\checkmark & \checkmark & \checkmark & \checkmark \\
	\bottomrule
\end{tabular}
\end{table}

\subsubsection{Weighted Networks}
\label{sec:weighted}

For the weighted case, we again consider $n \in \{240,600\}$ but 
focus on $K_\text{true} = 3$ only. Within-cluster link 
probabilities are set to $0.3$ and $0.6$, representing sparse and 
dense networks, respectively, with all within-cluster edge weights 
(denoted $w_\text{win}$) independently drawn from $\text{Unif}(0.5, 
1)$. The evaluation is conducted under varying between-cluster link 
probabilities: for $p_\text{win} = 0.3$, we set $p_\text{btw} = 0.1$ 
and $0.2$; for $p_\text{win} = 0.6$, we set $p_\text{btw} = 0.1$ and 
$0.5$. Both equal- and unequal-sized cluster configurations are 
examined, along with three distributions for between-cluster edge 
weights (denoted $w_\text{btw}$): no overlap with $w_\text{win}$ 
($w_\text{btw} \sim \text{Unif}(0,0.2)$), boundary overlap 
($w_\text{btw} \sim \text{Unif}(0.3,0.5)$), and substantial overlap 
($w_\text{btw} \sim \text{Unif}(0.5,0.7)$).

\subsubsection{Fully Connected Networks}
\label{sec:fully}

In addition, we conduct simulations for fully connected weighted 
networks, where every pair of nodes is linked. Although fully 
connected weighted networks are a special case of weighted networks, 
their performance notably differ from those observed in sparse 
weighted settings. For clarity, we therefore present results for this 
configuration separately. In this setting, only 
$w_\text{btw}$ and cluster size configurations are varied, with the 
same networks of $n \in \{240,600\}$ and $K_\text{true} = 3$. For 
$w_\text{btw}$, we again consider no overlap ($w_\text{btw} \sim 
\text{Unif}(0,0.2)$), boundary overlap ($w_\text{btw} \sim 
\text{Unif}(0.3,0.5)$), and substantial overlap ($w_\text{btw} \sim 
\text{Unif}(0.5,0.7)$). In addition, we examine even more substantial 
overlap ($w_\text{btw} \sim \text{Unif}(0.6,0.8)$) under both equal- 
and unequal-sized cluster configurations. Each scenario across all 
designs is replicated 200 times.

\subsection{Clustering Procedure}
\label{sec:clustering}

For each simulated network, clustering is performed using spectral 
clustering~\citep{vonluxburg2007atutorial}, with the number of 
clusters selected by maximizing the silhouette score. Given the 
adjacency matrix $\bm{W}$, we construct the normalized Laplacian,
$\bm{L}_{\text{norm}} = \bm{S}^{-1/2}(\bm{S} - \bm{W})\bm{S}^{-1/2}$,
where $\bm{S}$ is the diagonal strength matrix with entries $s_{ii} = 
\sum_{j=1}^n w_{ij}$. The first $K$ eigenvectors (corresponding to 
the smallest eigenvalues) of $\bm{L}_{\text{norm}}$ are extracted to 
form an $(n \times K)$ embedding matrix. Then, $k$-means clustering 
is applied to this embedding (after row-normalization), and the 
silhouette score is computed to 
evaluate clustering quality.

The number of clusters $K$ is selected by maximizing the silhouette 
score across candidate values $K \in \{2, \ldots, K_{\text{max}}\}$, 
where $K_{\text{max}}$ is set sufficiently large relative to 
$K_{\text{true}}$ to allow for overestimation; in our study, we set 
$K_{\text{max}} = 20$. In our data generation, all edge weights are 
drawn from $(0, 1)$, so we simply use $1$ minus edge weight as the 
distance metric. Finally clustering accuracy is evaluated using the 
adjusted Rand index~\citep[ARI,][]{hubert1985comparing}, which 
compares the cluster assignments with the true labels.

\subsection{Simulation Results}
\label{sec:sim_res}

\subsubsection{Unweighted Networks}
\label{sec:unweighted_res}

\begin{table}[tbp]
	\centering
	\renewcommand{\arraystretch}{1.5}
	\caption{Proportions of $K$ being correctly selected (out of $R 
		= 200$ simulation runs) under varying parameter settings: $n 
		\in \{240, 600\}$, $K_{\text{true}} 
		\in \{3, 8\}$, equal (EQ) or unequal (NE) cluster sizes.} 
	\label{tab:unweighted_n240_600}
	\small
	\setlength{\tabcolsep}{4.5pt}
	\begin{tabular}{c cccc cccc cccc cccc}
		\toprule
		& \multicolumn{8}{c}{$p_{\text{win}} = 0.3$} & 
		\multicolumn{8}{c}{$p_{\text{win}} = 0.4$} \\
		\cmidrule(lr){2-9} \cmidrule(lr){10-17} 
		& \multicolumn{4}{c}{$n = 240$} & \multicolumn{4}{c}{$n = 
		600$} & \multicolumn{4}{c}{$n = 240$} & 
		\multicolumn{4}{c}{$n = 600$} \\
		\cmidrule(lr){2-5} \cmidrule(lr){6-9} \cmidrule(lr){10-13} 
		\cmidrule(lr){14-17}
		& \multicolumn{2}{c}{$K_{\text{true}} = 3$} & 
		\multicolumn{2}{c}{$K_{\text{true}} = 8$} 
		& \multicolumn{2}{c}{$K_{\text{true}} = 3$} & 
		\multicolumn{2}{c}{$K_{\text{true}} = 8$}
		& \multicolumn{2}{c}{$K_{\text{true}} = 3$} & 
		\multicolumn{2}{c}{$K_{\text{true}} = 8$}
		& \multicolumn{2}{c}{$K_{\text{true}} = 3$} & 
		\multicolumn{2}{c}{$K_{\text{true}} = 8$} \\
		\cmidrule(lr){2-3} \cmidrule(lr){4-5} \cmidrule(lr){6-7} 
		\cmidrule(lr){8-9}
		\cmidrule(lr){10-11} \cmidrule(lr){12-13} 
		\cmidrule(lr){14-15} \cmidrule(lr){16-17}
		$p_{\text{btw}}$ & EQ & NE & EQ & NE & EQ & NE & EQ & NE 
		& EQ & NE & EQ & NE & EQ & NE & EQ & NE \\
		\midrule
		$0.05$ & $0.94$ & $0$ & $0.03$ & $0$ & $1$ & $0$ & $1$ & $0$ 
		& 
		$1$ & $0$ & $0.92$ & $0$ & $1$ & $0$ & $1$ & $0$ \\
		$0.1$ & $0.48$ & $0$ & $0$ & $0$ & $1$ & $0.17$ & $1$ & $0$ 
		& 
		$1$ & $0$ & $0.61$ & $0$ & $1$ & $0$ & $1$ & $0$ \\
		$0.15$ & $0$ & $0$ & $0$ & $0$ & $1$ & $0.44$ & $0.54$ & $0$ 
		& 
		$1$ & $0$ & $0$ & $0$ & $1$ & $0.13$ & $1$ & $0$ \\
		$0.2$ &   &   &   &   &   &   &   &   & $0.89$ & $0$ & $0$ & 
		$0$ & $1$ & $0.71$ & $1$ & $0$ \\
		$0.25$ &   &   &   &   &   &   &   &   & $0.11$ & $0$ & $0$ 
		& $0$ & $1$ & $0.16$ & $0.05$ & $0$ \\
		\midrule
		& \multicolumn{8}{c}{$p_{\text{win}} = 0.5$} & 
		\multicolumn{8}{c}{$p_{\text{win}} = 0.6$} \\
		\cmidrule(lr){2-9} \cmidrule(lr){10-17}
		& \multicolumn{4}{c}{$n = 240$} & \multicolumn{4}{c}{$n = 
		600$} 
		& \multicolumn{4}{c}{$n = 240$} & \multicolumn{4}{c}{$n = 
		600$} \\
		\cmidrule(lr){2-5} \cmidrule(lr){6-9} \cmidrule(lr){10-13} 
		\cmidrule(lr){14-17}
		& \multicolumn{2}{c}{$K_{\text{true}} = 3$} & 
		\multicolumn{2}{c}{$K_{\text{true}} = 8$} 
		& \multicolumn{2}{c}{$K_{\text{true}} = 3$} & 
		\multicolumn{2}{c}{$K_{\text{true}} = 8$}
		& \multicolumn{2}{c}{$K_{\text{true}} = 3$} & 
		\multicolumn{2}{c}{$K_{\text{true}} = 8$}
		& \multicolumn{2}{c}{$K_{\text{true}} = 3$} & 
		\multicolumn{2}{c}{$K_{\text{true}} = 8$} \\
		\cmidrule(lr){2-3} \cmidrule(lr){4-5} \cmidrule(lr){6-7} 
		\cmidrule(lr){8-9}
		\cmidrule(lr){10-11} \cmidrule(lr){12-13} 
		\cmidrule(lr){14-15} \cmidrule(lr){16-17}
		$p_{\text{btw}}$ & EQ & NE & EQ & NE & EQ & NE & EQ & NE 
		& EQ & NE & EQ & NE & EQ & NE & EQ & NE \\
		\midrule
		$0.1$ & $1$ & $0$ & $1$ & $0$ & $1$ & $0$ & $1$ & $0$ & $1$ 
		& $0$ & $1$ & $0$ & $1$ & $0$ & $1$ & $0$ \\ 
		$0.15$ & $1$ & $0.03$ & $1$ & $0$ & $1$ & $0$ & $1$ & $0$ 
		& $1$ & $0.01$ & $1$ & $0$ & $1$ & $0$ & $1$ & $0$ \\ 
		$0.2$ & $1$ & $0.02$ & $0.39$ & $0$ & $1$ & $0.1$ & $1$ & 
		$0$ & $1$ & $0.11$ & $1$ & $0$ & $1$ & $0$ & $1$ & $0$ \\ 
		$0.25$ & $1$ & $0.01$ & $0$ & $0$ & $1$ & $0.70$ & $1$ & $0$ 
		& $1$ & $0.17$ & $0.98$ & $0$ & $1$ & $0.02$ & $1$ & $0$ \\ 
		$0.3$ & $0.99$ & $0$ & $0$ & $0$ & $1$ & $0.59$ & $0.92$ & 
		$0$ & $1$ & $0.09$ & $0.13$ & $0$ & $1$ & $0.53$ & $1$ & $0$ 
		\\ 
		$0.35$ & $0.42$ & $0$ & $0$ & $0$ & $1$ & $0.08$ & $0$ & $0$ 
		& $1$ & $0.06$ & $0$ & $0$ & $1$ & $0.77$ & $1$ & $0$ \\ 
		$0.4$ & &  &  &  &  &  &  &  & $1$ & $0.04$ & $0.01$ & 
		$0.01$ & $1$ & $0.47$ & $0.74$ & $0$ \\ 
		$0.45$ &  &  &  & & &  &  & & $0.84$ & $0.02$ & $0.01$ & 
		$0.01$ & $1$ & $0.04$ & $0$ & $0$ \\ 
		\bottomrule
	\end{tabular}
\end{table}

\begin{figure}[tbh]
    \centering
    \includegraphics[width=0.9\textwidth]{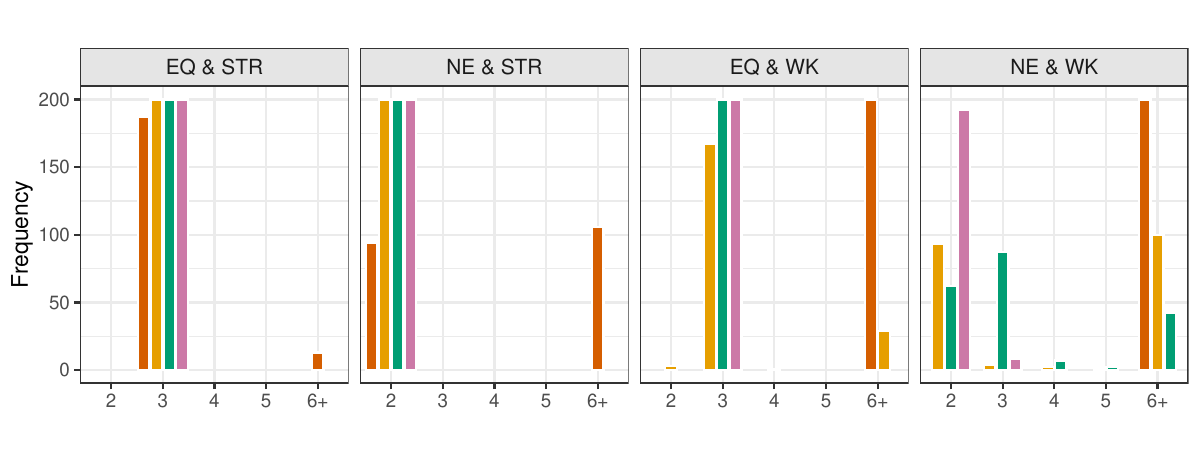}
 	\vspace{1em}
    \includegraphics[width=0.9\textwidth]{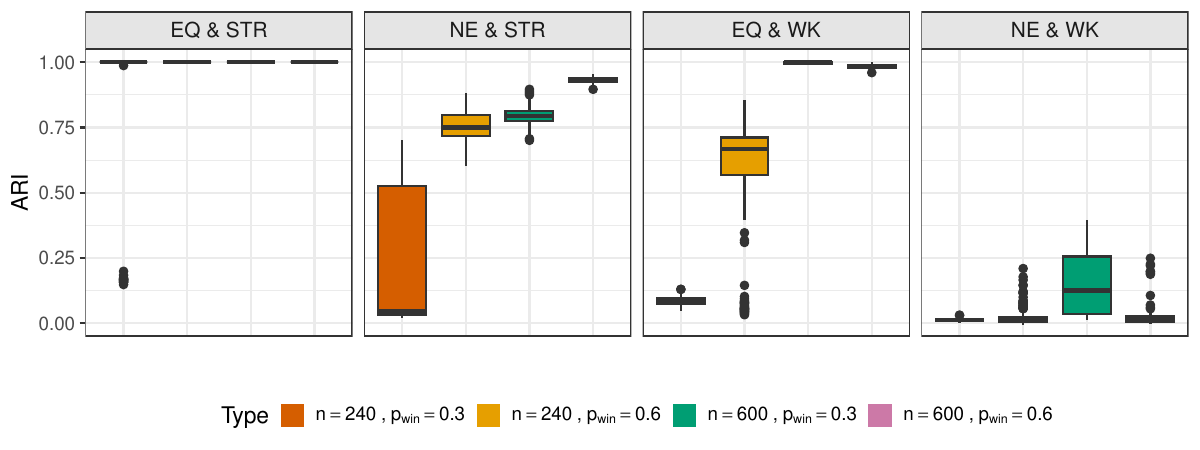}
  \caption{Example distributions of $K$ selection with 
  $K_\text{true} = 3$ across different 
  scenarios. For $p_\text{win} = 0.3$ and $0.6$, strongly (STR) 
  separated clusters correspond to $p_\text{btw} = 0.05$ and $0.1$, 
  respectively; weakly (WK) separated clusters have $p_\text{btw} = 
  0.15$ and $0.45$, respectively.}
  \label{fig:unweighted}
\end{figure}

Table~\ref{tab:unweighted_n240_600} summarizes the proportion of 
simulation runs in which the true $K$ was correctly selected under 
various simulation settings. Overall, when clusters are well 
separated (i.e., $p_{\text{win}} > p_{\text{btw}}$ significantly), 
the silhouette score consistently identifies the correct number of 
clusters, regardless of network size or number of clusters. However, 
when one cluster is much larger than the others, the silhouette 
score often underestimates $K_{\text{true}}$, since merging smaller 
clusters tends to inflate the score; see the two NE panels in the 
top row of Figure~\ref{fig:unweighted} for examples. In sparse 
networks, the silhouette score may instead overestimate 
$K_{\text{true}}$ due to the formation of singletons; see the red
scenario under the two NE panels in the top row of 
Figure~\ref{fig:unweighted} for examples. In such cases, incorrect 
selection of $K$ is largely attributed to these 
network configurations.

Notably, with balanced cluster sizes, correct selection of $K$ is 
more strongly associated with accurate clustering, particularly in 
larger networks, as evidenced by the comparison between histograms 
and the corresponding ARI box plots in Figure~\ref{fig:unweighted}. 
When network size is relatively small but 
dense, clustering performance is moderate rather than perfect, with 
ARI averaging around $0.68$ even though the correct $K$ is selected 
in roughly $85\%$ of simulation runs. A closer examination of the 
case with $n = 240$, $p_{\text{win}} = 0.6$, and $p_{\text{btw}} = 
0.15$ (WK) illustrates this point. Although only about $8\%$ of nodes 
are misclassified, they are distributed across all clusters, which 
substantially reduces average ARI values. The outliers observed in 
the box plot (the orange scenario under the ``EQ \& WK'' panel in the 
bottom row of Figure~\ref{fig:unweighted}) correspond 
to the instances where $K$ is incorrectly selected. In addition, 
when networks are small and sparse, the silhouette score tends to 
severely overestimate $K$ due to the existence of singletons, 
consequently leading to extremely poor clustering performance.

In contrast, when cluster sizes are imbalanced, the silhouette score 
often favors a smaller~$K$ (e.g., selecting $K = 2$ when 
$K_{\text{true}} = 3$), as merging the 
two smaller clusters usually tends to increase the score. When the 
cluster structure is strongly defined, merging smaller clusters may 
influence the dominant cluster, but only to a limited extent, so 
relatively high ARI values can still be achieved. This effect is 
further attenuated in larger and denser networks, where the dominant 
cluster becomes even more dominant, thereby mitigating the impact of 
selecting an incorrect $K$. However, in smaller and sparser networks, 
clustering performance is highly variable and generally poor. When 
clusters are weakly separated, the silhouette score almost never 
identifies the correct $K$, and clustering performance deteriorates 
substantially. Similar patterns are observed for larger values of 
$K_{\text{true}}$, with additional examples and discussion provided 
in Appendix~\ref{app:hist_ari_k_8}.


Importantly, even when $K$ is sometimes correctly 
selected under imbalanced cluster sizes, the resulting partitions 
are of poor quality, as reflected in low ARI values. For instance, 
when $n = 600$, $K_{\text{true}} = 3$, $p_{\text{win}} = 0.4$, and 
$p_{\text{btw}} = 0.2$ (in Table~\ref{tab:unweighted_n240_600}), the 
proportion of correctly selecting $K$ under NE is $0.71$, but the 
median and interquartile ranges of ARI are $0.29$ and $0.15$, 
respectively. To highlight this behavior, 
Appendix~\ref{app:ari_ne_ge0.5} presents box plots 
for all NE cases with correct selection proportions exceeding $0.5$ 
in Table~\ref{tab:unweighted_n240_600} for illustration. In contrast, 
when cluster sizes are equal, the silhouette score more reliably 
recovers the correct~$K$. Its performance improves with larger 
network size but deteriorates when cluster number increases and 
$p_{\text{btw}}$ approaches $p_{\text{win}}$ (which results in weaker 
cluster separation).

When the network is relatively small and sparse while 
the number of clusters is large (e.g., $n = 240$, $K_{\text{true}} = 
8$, $p_{\text{win}} = 0.3$, $p_{\text{btw}} = 0.05$ in 
Table~\ref{tab:unweighted_n240_600}), the silhouette score can 
perform poorly (proportion of selecting the correct $K = 0.03$) even 
when cluster sizes are balanced and 
the cluster separation appears clear. This is because the 
combination of a modest sample size and many clusters yields small 
clusters, and a low $p_{\text{win}}$ induces high within-cluster 
sparsity. These factors together produce an extremely weak community 
structure for the silhouette score to capture effectively.

\begin{table}[tbh]
	\centering
	\renewcommand{\arraystretch}{1.25}
	\caption{Proportions of $K$ being correctly selected (out of $R 
	= 200$ simulation runs) when $K_\text{true} = 3$ under varying 
	parameter settings: $n \in \{240, 600\}$, cluster size being 
	equal (EQ) or unequal (NE), and one pair of clusters is selected 
	to be weakly separated with $\tilde{p}_{\text{btw}}$ specified in 
	the table. All other between-cluster probabilities 
	${p}_\text{btw}$ are fixed at $0.05$.} 
	\label{tab:unweighted_one_pair}
	\setlength{\tabcolsep}{12pt}
	\small
	\begin{tabular}{c cc cc cc cc}
		\toprule
		& \multicolumn{4}{c}{$p_{\text{win}} = 0.3$} & 
		\multicolumn{4}{c}{$p_{\text{win}} = 0.6$} \\
		\cmidrule(lr){2-5} \cmidrule(lr){6-9}
		& \multicolumn{2}{c}{$n = 240$} & \multicolumn{2}{c}{$n = 
		600$} 
		& \multicolumn{2}{c}{$n = 240$} & \multicolumn{2}{c}{$n = 
		600$} \\
		\cmidrule(lr){2-3} \cmidrule(lr){4-5} 
		\cmidrule(lr){6-7} \cmidrule(lr){8-9}
		$\tilde{p}_{\text{btw}}$ & EQ & NE & EQ & NE & EQ & NE & EQ & 
		NE \\
		\midrule
		$0.1$  & $0.74$ & $0$ & $1$ & $0$ & & & & \\
		$0.15$ & $0.02$ & $0$ & $0.17$ & $0$ & & & & \\
		$0.2$  & $0$ & $0$ & $0$ & $0.13$ & & & & \\
		$0.3$  & & & & & $0.96$ & $0$ & $1$ & $0$ \\
		$0.35$ & & & & & $0.43$ & $0$ & $1$ & $0$ \\
		$0.4$  & & & & & $0$ & $0$ & $1$ & $0$ \\
		$0.45$ & & & & & $0$ & $0$ & $0.4$ & $0$ \\
		\bottomrule
	\end{tabular}
\end{table}

\begin{figure}[tbh]
	\centering
	\includegraphics[width=0.9\textwidth]{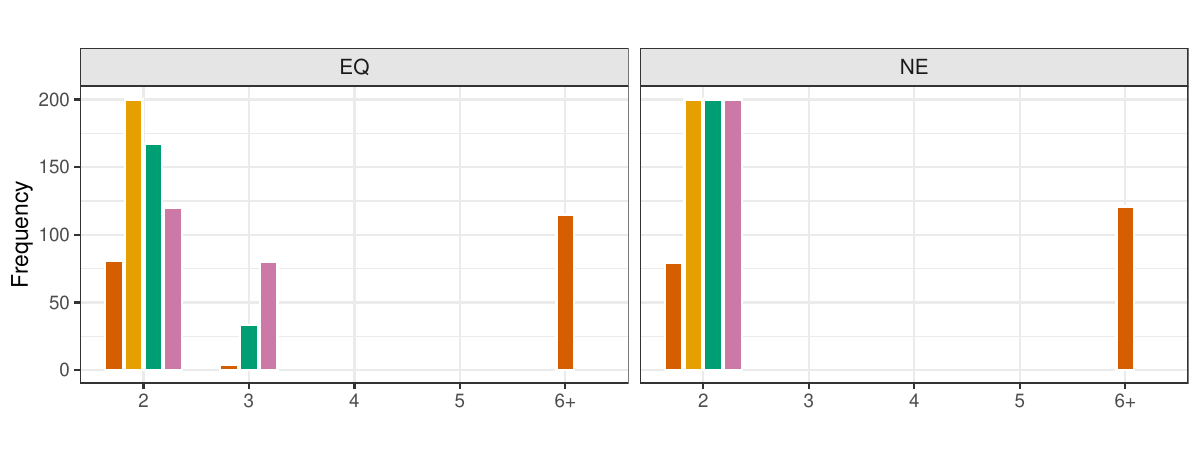}
	\vspace{1em}
	\includegraphics[width=0.9\textwidth]{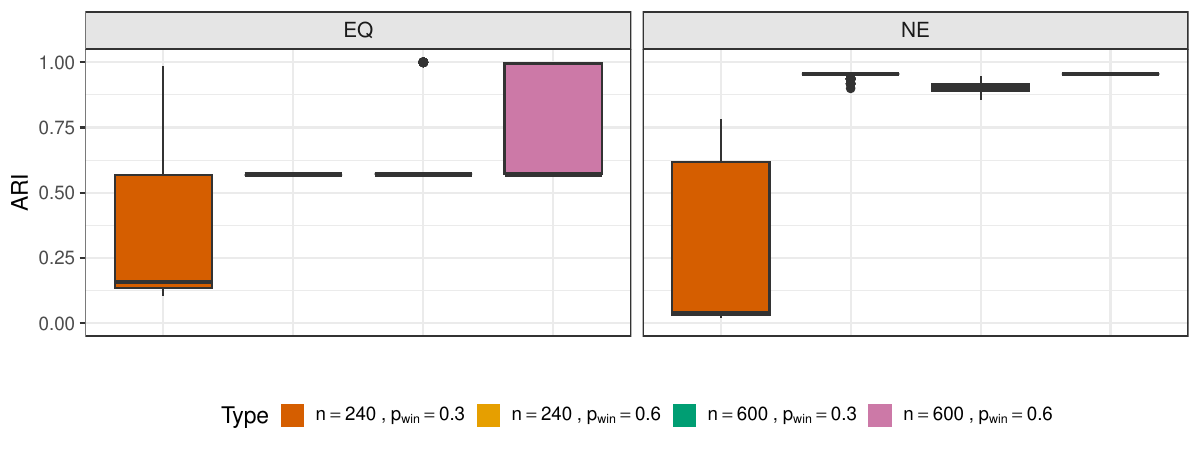}
	\caption{Example distributions of $K$ selection and ARI box plots 
		with $K_\text{true} = 3$ under scenarios with a pair of 
		weakly 
		separated clusters. For $p_\text{win} = 0.3$, the weakly 
		separated 
		pair has $\tilde{p}_\text{btw} = 0.15$ while the remaining 
		pairs have 
		$p_\text{btw} = 0.05$. For $p_\text{win} = 0.6$, the weakly 
		separated 
		pair has $\tilde{p}_\text{btw} = 0.45$ and the others have 
		$p_\text{btw} = 
		0.05$.}
	\label{fig:one_pair_weak}
\end{figure}

For $K_{\text{true}} = 3$, when either a pair of equally sized 
clusters are weakly separated or the two smaller clusters in an 
imbalanced setting are weakly separated, the correct $K$ is rarely 
selected, regardless of cluster size balance, as shown in 
Table~\ref{tab:unweighted_one_pair} and 
Figure~\ref{fig:one_pair_weak}. In small and sparse networks, the 
silhouette score often overestimates $K$ due to the emergence of 
singletons, whereas in most other settings it underestimates by 
selecting $K = 2$. For large or dense networks, clustering 
performance under imbalanced sizes is generally above average because 
of the dominance of the large cluster, minimally affected by 
separation strength. In contrast, when cluster sizes are balanced but 
separation is weak, ARI values remain unsatisfactory, though not 
disastrous, primarily due to node misclassification between weakly 
separated clusters.

\subsection{Weighted Networks}
\label{sec:weighted_res}

\begin{figure}[tbh]
	\centering
		\includegraphics[width=0.9\textwidth]{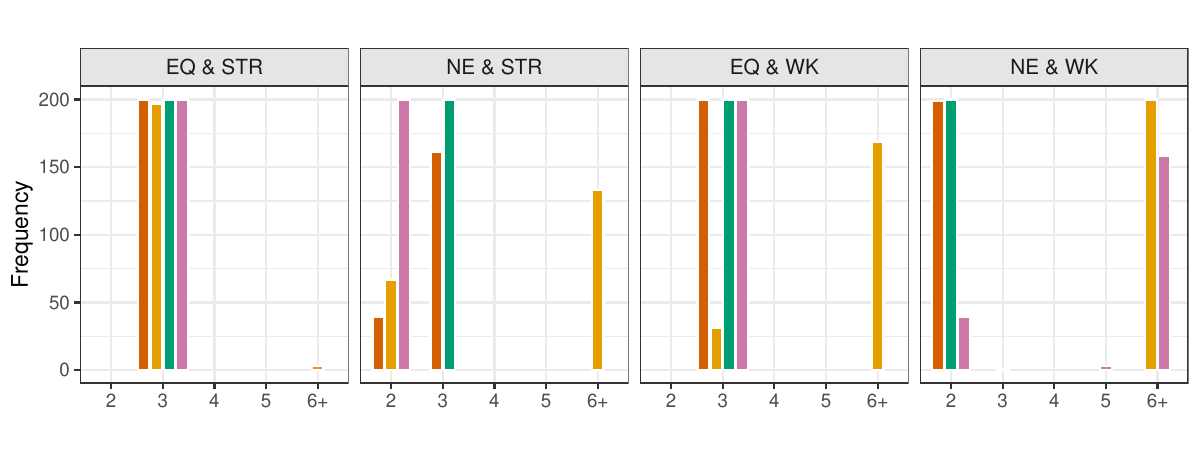}
		\vspace{1em}
		\includegraphics[width=0.9\textwidth]{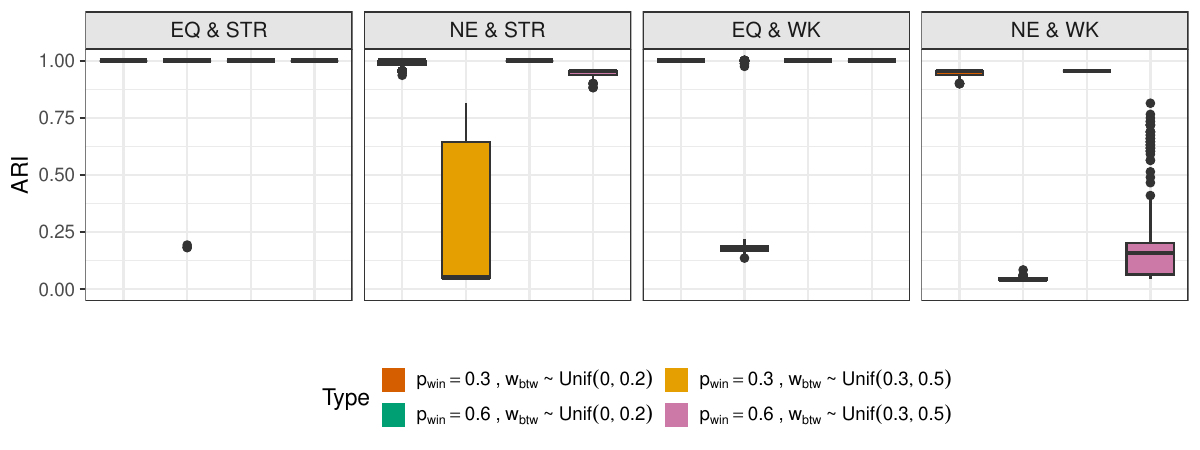}
		
	\caption{Example distributions of $K$ selection and ARI box plots 
	for weighted networks with $K_\text{true} = 3$ and $n = 240$, 
	where within-cluster weights are sampled from $\text{Unif}(0.5, 
	1)$. For $p_\text{win} = 0.3$ and $0.6$, strongly separated (STR)
	clusters correspond to $p_\text{btw} = 0.1$, while weakly 
	separated (WK) clusters correspond to $p_\text{btw} = 0.2$ and 
	$0.5$, 
	respectively.}
	\label{fig:weighted_n240}
\end{figure}

Figure~\ref{fig:weighted_n240} presents the proportion of correct $K$ 
selections along with ARI-based clustering performance for weighted 
networks with $K_\text{true} = 3$ and $n = 240$. When clusters are of 
equal size, the silhouette score generally recovers the correct $K$ 
with nearly perfect ARI, regardless of within- and between-cluster 
connectivity or the strength of between-cluster edge weights; see 
the histograms and ARI box plots in the ``EQ \& STR'' and ``EQ \& 
WK'' panels of Figure~\ref{fig:weighted_n240}. The only 
exception arises when networks are sparse and clusters are weakly 
separated ($p_\text{win} = 0.3$, $p_\text{btw} = 0.2$) with 
between-cluster edge weights nearly overlapping the within-cluster 
distribution ($w_\text{win} \sim \text{Unif}(0.5, 1)$ and 
$w_\text{btw} \sim \text{Unif}(0.3,0.5)$); see the orange scenario 
under the ``EQ \& WK'' panel of Figure~\ref{fig:weighted_n240}. In 
this case, the silhouette score tends to overestimate $K$ because the 
weak separation in connectivity is further exacerbated by close edge 
weights that blur cluster boundaries, leading to many singletons in 
sparse networks. As a result, ARI values remain consistently low due 
to incorrect $K$. The few outliers near~$1$ in the box plot 
correspond to rare cases where the correct $K$ is selected. Overall, 
these patterns closely mirror those observed in unweighted networks.

However, it is notable that in the same sparse networks with weakly 
separated clusters ($p_\text{win} = 0.3$, $p_\text{btw} = 0.2$), the 
silhouette score can recover the correct $K$ when between-cluster 
edge weights are substantially smaller than within-cluster ones 
($w_\text{win} \sim \text{Unif}(0.5,1)$ and $w_\text{btw} \sim 
\text{Unif}(0,0.2)$; see the red scenario under the ``EQ \& STR'' and 
``EQ \& WK'' panels in the top row of 
Figure~\ref{fig:weighted_n240}. In this setting, between-cluster 
edges receive small weights despite similar connectivity patterns, 
limiting their influence on the inferred cluster structure. 
Consequently, the effect of sparsity is mitigated, 
singletons do not appear, and clustering performance is notably 
improved compared to the unweighted counterpart.

For imbalanced clusters, the ability of the silhouette score to 
correctly identify $K$ depends on both the degree of separation and 
the distribution of between-cluster edge weights. When clusters are 
strongly separated ($p_\text{win} = 0.3$ or $0.6$ with 
$p_\text{btw} = 0.1$), the silhouette score consistently selects the 
correct $K$ with nearly perfect ARI (For $p_\text{win} = 0.3$, a 
small fraction of runs underestimate $K$, though the resulting ARI 
remains high overall), provided that between-cluster edge weights are 
substantially smaller 
than within-cluster ones ($w_\text{win} \sim \text{Unif}(0.5, 1)$ and 
$w_\text{btw} \sim \text{Unif}(0, 0.2)$); see the red and green 
panels under the ``NE \& STR'' panel of 
Figure~\ref{fig:weighted_n240}. In this setting, the 
networks already contain few between-cluster edges, and the small 
between-cluster weights further sharpen cluster boundaries. As a 
result, merging smaller clusters no longer increases the silhouette 
score, leading to accurate recovery of the true $K$.

However, when between-cluster edge weights are closer to 
within-cluster values ($w_\text{win} \sim \text{Unif}(0.5,1)$ and 
$w_\text{btw} \sim \text{Unif}(0.3,0.5)$), the silhouette score fails 
to recover the correct $K$; see the orange and pink scenarios under 
the ``NE \& STR'' panel in the top row of 
Figure~\ref{fig:weighted_n240}. Specifically, in denser 
networks ($p_\text{win} = 0.6$, $p_\text{btw} = 0.1$), the silhouette 
score tends to underestimate $K$ by merging the two smaller clusters. 
Nevertheless, the ARI median remains high because the largest cluster 
dominates the evaluation. Compared to unweighted networks, the ARI 
distribution in this setting shows higher medians and smaller 
variation, suggesting that incorporating edge weights can reduce the 
effective strength of between-cluster ties. This observation 
highlights the importance of incorporating weights into clustering 
when they are available. However, in sparser networks ($p_\text{win} 
= 0.3$, $p_\text{btw} = 0.1$), the silhouette score often 
overestimates $K$ due to the frequent appearance of singletons, with 
even the largest cluster fragmented into smaller pieces. Ambiguous 
within- and between-cluster edge weights further obscure cluster 
boundaries, leading to clustering performance that shows no 
improvement over the unweighted counterparts.

Finally, when imbalanced clusters are weakly separated ($p_\text{win} 
= 0.3$ with $p_\text{btw} = 0.2$ or $p_\text{win} = 0.6$ with 
$p_\text{btw} = 0.5$), the silhouette score consistently fails to 
choose the correct $K$; see the ``NE \& WK'' panel in the top row of 
Figure~\ref{fig:weighted_n240}. With small between-cluster weights 
($w_\text{btw} \sim \text{Unif}(0,0.2)$), the silhouette score 
typically underestimates $K$ by merging the two smaller clusters, but 
the dominance of the largest cluster still drives relatively high ARI 
values. In contrast, with larger between-cluster weights 
($w_\text{btw} \sim \text{Unif}(0.3,0.5)$), the silhouette score 
tends to overestimate $K$ in sparse networks ($p_\text{win} = 0.3$, 
$p_\text{btw} = 0.2$), leading to poor clustering performance 
reflected in low ARI values. Notably, in denser networks 
($p_\text{win} = 0.6$, $p_\text{btw} = 0.5$), $K$ selection exhibits 
much greater variability. Most runs break the largest cluster into 
smaller pieces, producing ARI values near $0.2$, while occasional 
runs merge the two smaller clusters and mostly preserve the 
dominant one, yielding ``better'' outliers in the ARI 
distribution.

We conduct an additional analysis for larger networks ($n = 600$) and 
observe similar patterns, although the increased sample size 
generally mitigates the effects of weak cluster separation and 
ambiguous between-cluster edge weights (relative to within-cluster 
edges). For this reason, the results for this analysis are omitted.

\subsection{Fully Connected Networks}
\label{sec:fullyconn_res}

For fully connected networks, we focus only on weighted networks, 
with within-cluster weights sampled from $\text{Unif}(0.5,1)$. We 
observe that when between-cluster weights deviate substantially from 
within-cluster weights ($w_\text{btw} \sim \text{Unif}(0,0.2)$) or 
are close but non-overlapping ($w_\text{btw} \sim 
\text{Unif}(0.3, 0.5)$), the correct $K$ is consistently 
selected and ARI values are always perfect, without variation across 
network size and cluster size distributions (whether equal-sized or 
dominated by a large cluster). Consequently, we account for 
between-cluster weights sampled from $\text{Unif}(0.5, 0.7)$ and 
$\text{Unif}(0.6, 0.8)$, which respectively represent different 
degrees of overlap with within-cluster weights and thus illustrate 
varying levels of separation.

\begin{figure}[tbh]
		\centering
		\includegraphics[width=0.9\textwidth]{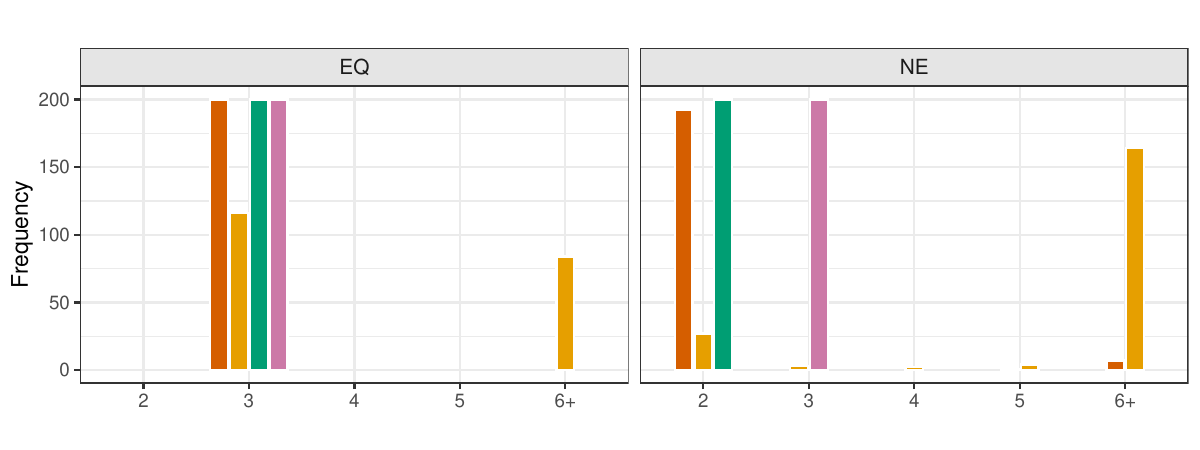}
		\vspace{1em}
		\includegraphics[ width=0.9\textwidth]{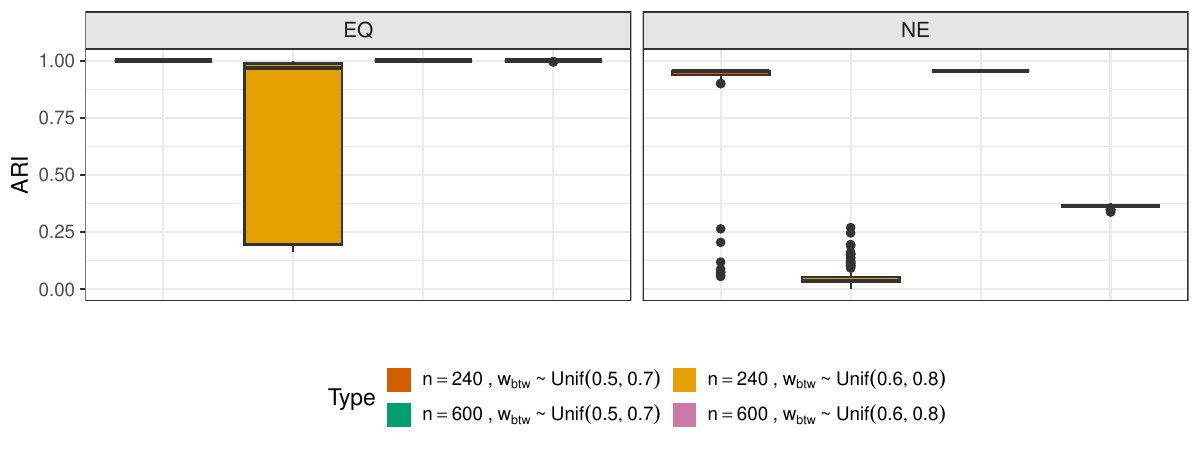}
	\caption{Example distributions of $K$ selection and ARI box 
	plots for fully connected, weighted networks with $K_\text{true} 
	= 3$, where within-cluster weights are sampled from 
	$\text{Unif}(0.5, 1)$. Between-cluster weights are sampled from 
	$\text{Unif}(0.5, 0.7)$ and $\text{Unif}(0.6, 0.8)$ to 
	illustrate different levels of weight separation.}
	\label{fig:fully}
\end{figure}

Figure~\ref{fig:fully} shows the proportion of correctly selected 
$K$ along with the corresponding ARI box plots. For equal-sized 
clusters, the correct $K$ is consistently selected provided that 
between-cluster weights do not strongly overlap with within-cluster 
weights ($w_\text{btw} \sim \text{Unif}(0.5,0.7)$). In these cases, 
clustering performance is nearly perfect, with ARI values close to 
$1$ alongside minimal variability. However, when between-cluster 
weights overlap substantially with within-cluster weights 
($w_\text{btw} \sim \text{Unif}(0.6,0.8)$), performance depends on 
network size. In large networks, the correct $K$ is still recovered 
reliably, but in smaller networks the silhouette score selects the 
correct $K$ only about $58\%$ of the time, with the remainder 
tending to overestimate $K$; see the ``EQ'' panel in the top row of 
Figure~\ref{fig:fully}. In this setting, the silhouette 
score is inflated when each target cluster is fragmented into 
several small clusters, primarily due to substantial overlap in 
between-cluster weights. Consequently, these incorrect selections of 
$K$ introduce large variability in ARI, with an interquartile range 
(IQR) of $0.79$, even though the overall median remains high (around 
$0.97$).

For imbalanced clusters, the silhouette score often underestimates 
$K$ by merging smaller clusters when the between-cluster weights do 
not heavily overlap with the within-cluster weights ($w_\text{btw} 
\sim \text{Unif}(0.5,0.7)$); see the ``NE'' panel in 
the top panel of Figure~\ref{fig:fully}. However, the overall 
clustering accuracy, as reflected in ARI values, is good, especially 
for large networks. For smaller networks, we observe outliers in the 
ARI distribution, which correspond to the few cases with 
overestimated $K$, where the largest cluster is broken into several 
smaller clusters due to weight ambiguity.

As between-cluster weights increasingly overlap with within-cluster 
weights ($w_\text{btw} \sim \text{Unif}(0.6,0.8)$), the silhouette 
score rarely selects the true $K$. In small networks, both 
overestimation and underestimation occur. Especially when $K$ is 
underestimated ($K = 2$), the method does not simply merge the two 
smaller clusters; instead, it fragments the large cluster and 
regroups its nodes with those from the smaller cluster, yielding 
extremely poor clustering accuracy. In larger networks, the 
silhouette score often identifies the correct $K$, but ARI remains 
poor. Further inspection shows the large cluster is again split into 
several sub-clusters that are then merged with nodes from the two 
smaller clusters. This highlights a key caveat: correctly estimating 
$K$ does not always guarantee strong clustering performance, 
particularly under cluster size imbalance cases.

\section{Airline Reachability Network Analysis}
\label{sec:air}

In this section, we analyze the Airline Reachability 
Network~\citep[ARN,][]{frey2007clustering}, using the silhouette 
score to estimate the number of clusters. The dataset includes $n = 
456$ cities in the United States (including Hawaii and Alaska) and 
Canada. Each directed edge is weighted by the negative estimated 
airline travel time between cities, including stopover delays. 
Although round-trip service is common, the ARN is asymmetric due to 
two factors: (1) prevailing wind conditions, and (2) the omission of 
routes whose total travel time exceeds $48$ hours.

We apply standard spectral clustering to this network; however, the 
method does not directly accommodate negative weights or asymmetry. 
To retain the essential topology of ARN, we preprocess the network as 
follows. First, we scale all edge weights to the $[0,1]$ range:
\[w_{ij} \leftarrow \frac{w_{ij} - 
\min(\bm{w})}{\text{max}(\bm{w})-\text{min}(\bm{w})},\]
where $\bm{w} = \{w_{ij}: 1 \le i \neq j \le n\}$ denotes all entries 
in the weighted adjacency matrix.
Next, since our analysis focuses on mutually reachable city pairs, we 
enforce symmetry by averaging the weights of reciprocal edges when 
both directions are present: \[w_{ij} = w_{ji} \leftarrow 
\frac{w_{ij} + w_{ji}}{2}, \qquad \text{if } w_{ij} > 0 \text{ and } 
w_{ji} > 0.\] 
After these preprocessing steps, the resulting weighted, undirected 
network contains $34{,}011$ edges among $456$ nodes, corresponding 
to approximately $67\%$ sparsity.

\begin{table}[tbh]
	\centering
	\small
	\renewcommand{\arraystretch}{1.2}
	\setlength{\tabcolsep}{12pt}
	\caption{Cluster sizes (in parentheses) and within- and 
		between-cluster densities (percent) for ARN.}
	\label{tab:real_cluster}
	\begin{tabular}{rccccc}
		\toprule
		Cluster & 1 $(141)$ & 2 $(103)$ & 3 $(96)$ & 4 $(76)$ & 5 
		$(40)$ \\ 
		\midrule
		1 & 60\% & 29\% & 34\% & 23\% & 14\% \\ 
		2 &  & 60\% & 30\% & 21\% & 27\% \\ 
		3 &  &  & 45\% & 26\% & 19\% \\ 
		4 &  &  &  & 51\% & 19\% \\ 
		5 &  &  &  &  & 52\% \\ 
		\bottomrule
	\end{tabular}
\end{table}

\begin{figure}[tbp]
	\centering
	\includegraphics[width = 0.65\textwidth]{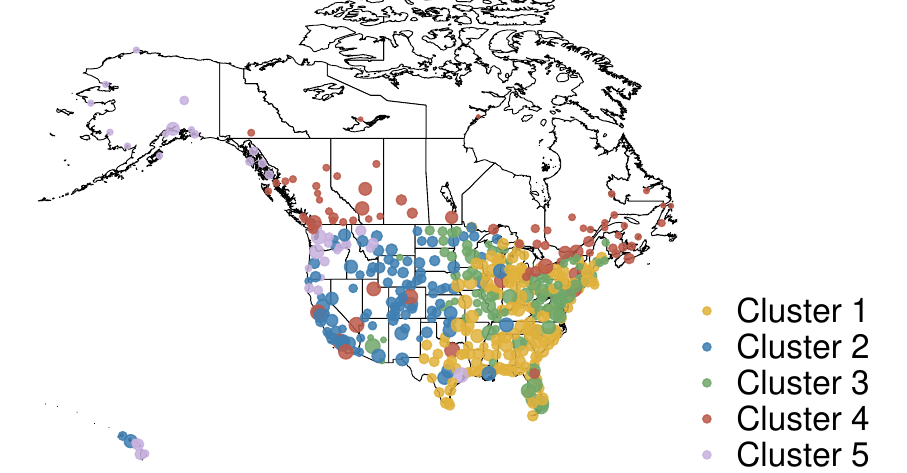}
	\caption{Clustering results displayed on the geographic map of 
	North America, with each city colored according to its assigned 
	cluster. Larger circle indicates a higher node degree.}
	\label{fig:map}
\end{figure}

We apply spectral clustering to ARN and select the number of clusters 
by maximizing the silhouette score over candidate values $K \in \{2, 
3, \ldots, 20\}$, yielding an optimal choice of $K = 5$. 
Table~\ref{tab:real_cluster} reports the within- and between-cluster 
densities. Overall, clusters exhibit clearly stronger internal 
connectivity than external connectivity in the network. Additionally, 
Figure~\ref{fig:map} displays the clustering results on a geographic 
map of North America, using location and metropolitan population 
information from \citet{benson2016higher}. Cluster memberships are 
indicated by color, and node sizes are proportional to the strength 
of 
the nodes (i.e., the sum of incident edge weights).

Overall, the detected clusters align strongly with geographic regions 
and major hub-and-spoke structures in North American air travel, 
reflecting airline routing economics and market segmentation. Cluster 
1 is the largest group, consisting primarily of cities in the 
Midwest and Eastern United States. While Long Island MacArthur, NY 
has the largest metropolitan population in this cluster, the most 
highly connected airports are Kansas City, MO and Cincinnati, OH, 
suggesting that Cluster 1 is characterized by mid-sized cities that 
serve mainly as regional connectors. Cluster 3 overlaps 
geographically with Cluster 1 but contains more 
major metropolitan airports and prominent internal hubs, including 
Washington, DC, Philadelphia, PA, and Minneapolis--St Paul, MN. 
These cities display notably higher degrees and function as key 
connectors in the U.S.\ airline network, in contrast to the more 
regionally oriented airports in Cluster 1. Cluster 2 contains cities 
along the West Coast and throughout the Mountain region, with a 
concentration of large metropolitan airports in California, such as 
Burbank, CA and Oakland, CA, respectively serving the Los Angeles 
and San Francisco areas. Cluster 4 comprises a relatively 
small number of major international hubs in the United States, such 
as Los Angeles, CA, New York, NY, Chicago, IL, and Dallas/Fort 
Worth, TX, alongside several cities in New England. This cluster 
also extends into Canada, with Toronto, ON and Vancouver, BC acting 
as key hubs and connecting most Canadian cities outside the West 
Coast. Cluster 5, in contrast, primarily consists of West Coast 
cities, particularly those in northern California, Oregon, 
Washington and Canada, along with all major Alaskan cities, with 
Seattle/Tacoma, WA serving as a central hub. Both Clusters 4 and 5 
also include cities located in Hawaii.

\section{Discussion}
\label{sec:discuss}

This study provides a comprehensive empirical evaluation of the 
silhouette score for selecting the number of clusters in network 
data. Although the silhouette score has been widely studied and 
recently extended in a range of clustering settings, its behavior in 
network clustering, particularly its success and failure modes under 
controlled variation in network size, community separation, and 
community-size imbalance, has not been sufficiently examined.

Our analysis highlights how these network characteristics shape 
silhouette-based selection across unweighted, weighted, and fully 
connected networks. We find that the silhouette score performs well 
when networks exhibit clear community structure and relatively 
balanced cluster sizes, but its performance deteriorates 
substantially as cluster imbalance increases, especially in sparse 
networks. Incorporating edge weights can improve robustness to 
sparsity when within- and between-community edge weights are well 
separated or only moderately overlapping; however, performance 
degrades when these distributions overlap substantially. The 
silhouette score also performs reliably in fully connected networks, 
although the combination of cluster imbalance and overlapping edge 
weights can still bias cluster number selection, often leading to 
underestimation of the true number of clusters. Notably, in some 
imbalanced settings, the silhouette score may correctly identify the 
true number of clusters while producing inaccurate cluster 
assignments.

Together, these findings provide practical guidance on when 
silhouette-based methods are reliable in network clustering and when 
caution is warranted, particularly in heterogeneous networks or 
systems dominated by a single large community. In addition, the 
airline reachability network application demonstrates that 
silhouette-based clustering can recover meaningful structure aligned 
with geography and market segmentation, suggesting its practical 
utility in real-world network data.

Despite offering critical insights, this study has several 
limitations. First, we employ the SBM to define cluster structures, 
which, while widely used, may not fully 
capture the heterogeneity and overlapping community organization 
present in real-world systems. Besides, our simulation study primarily
focuses on edge weights from the $[0,1]$ range without considering 
extreme or heavy-tailed weight distributions. Additionally, our 
analysis does not extensively address complex or non-convex cluster 
geometries. A simple example illustrating potential effects of 
non-convexity is provided in Appendix~\ref{app:convex}, although a 
systematic study remains for future work. Other future research may 
extend these analyses to more flexible 
network models, including degree-corrected or overlapping SBMs, or 
even to dynamic networks where community structure changes over 
time. Another promising direction is to develop adjusted or weighted 
variants of the silhouette score that explicitly account for cluster 
size imbalance, sparsity, and heterogeneous edge-weight 
distributions. Such methodological extensions could enhance its 
utility for large-scale applications in biology and social science, 
where data often exhibit both modular and hierarchical organization.









\bibliographystyle{chicago}
\bibliography{refs}

\appendix

\section{Additional ARI Box Plots for Imbalanced Cluster Sizes}
\label{app:ari_ne_ge0.5}

\begin{figure}[tbh]
  \centering
  \includegraphics[width=0.9\textwidth]{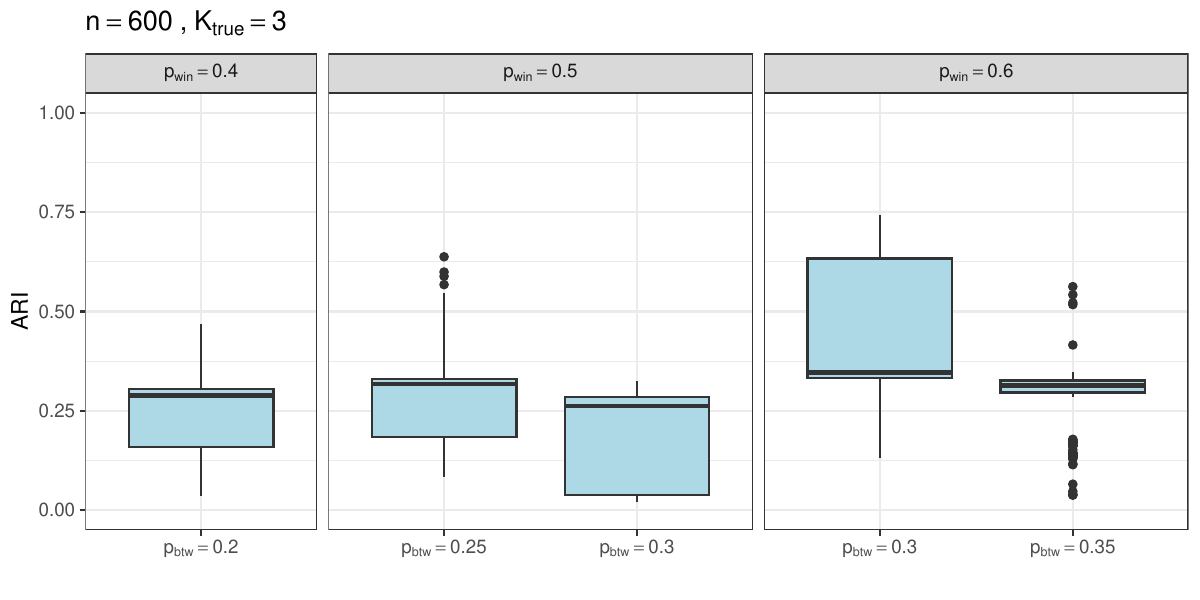}
  \caption{ARI box plots for imbalanced cluster sized with $>50\%$ 
    correct $K$ selection in 
    Table~\ref{tab:unweighted_n240_600}.}
  \label{fig:ari_ne_ge0.5}
\end{figure}

As shown in Table~\ref{tab:unweighted_n240_600}, the proportion of 
correct $K$ selection for imbalanced cluster sizes using the 
silhouette score is generally low. However, there are five specific 
scenarios in which the proportion of correct $K$ selections exceeds 
$50\%$. A common pattern is that all these scenarios occur with a 
larger network size ($n = 600$) and a smaller number of clusters 
($K_{\text{true}} = 3$). As shown in Figure~\ref{fig:ari_ne_ge0.5}, 
the ARI medians for all five scenarios are low (between $0.25$ and 
$0.4$) with varying variability, indicating that even when the 
silhouette score occasionally selects the correct $K$ for imbalanced 
cluster sizes, the resulting clustering performance is 
unsatisfactory.

A further investigation shows that, when $n$ is large and 
$K_{\text{true}}$ is small, the large cluster (under imbalanced 
cluster sizes) is more dominant and can usually be well identified. 
When $p_{\text{btw}}$ is relatively close to $p_{\text{win}}$, the 
silhouette score can be inflated when the ``boundary nodes'' of the 
small clusters are merged into the large one, while the two 
remaining small clusters persist. In this case, the silhouette score 
may still select the correct number of clusters with higher 
probability, but the resulting clustering significantly deviates 
from the truth, leading to low ARI values.

\section{Histograms and ARI Box Plots for $K_{\text{true}} = 8$}
\label{app:hist_ari_k_8}

\begin{figure}[tbh]
  \centering
  \includegraphics[width = 0.9\textwidth]{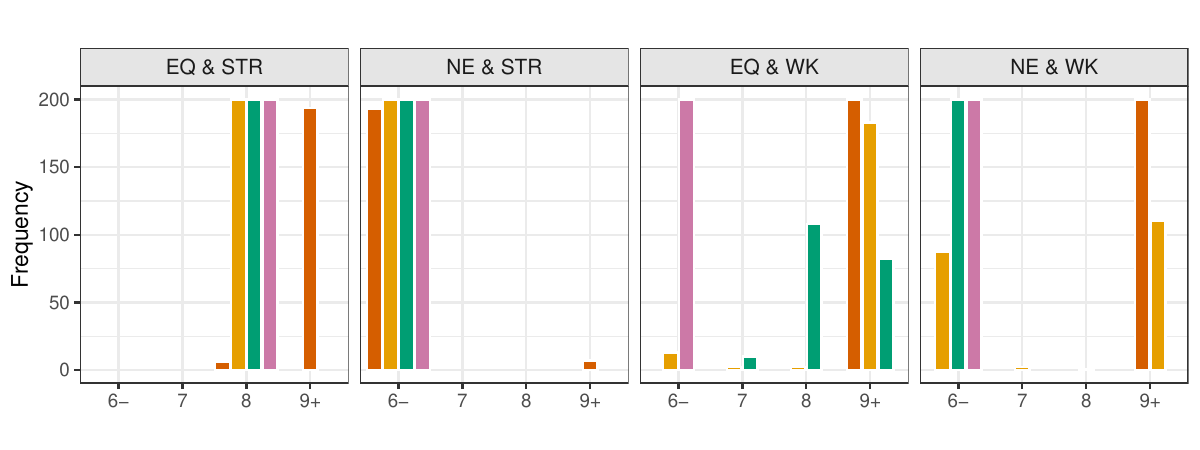}
  \vspace{1em}
  \includegraphics[width = 0.9\textwidth]{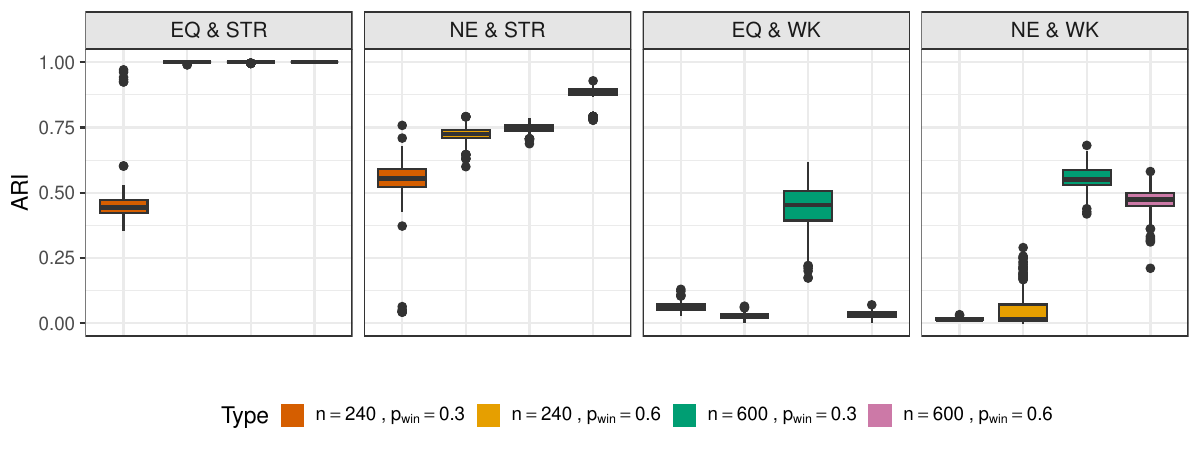}
    \caption{Example distributions of $K$ selection and ARI box plots 
with $K_\text{true} = 8$ across different scenarios.  For 
$p_\text{win} = 0.3$ and $0.6$, strongly (STR) separated clusters 
correspond to $p_\text{btw} = 0.05$ and $0.1$, respectively; weakly 
(WK) separated clusters have $p_\text{btw} = 0.15$ and $0.45$, 
respectively.}
\label{fig:all_weak_k_8}
\end{figure}

As shown in Figure~\ref{fig:all_weak_k_8}, when clusters are weakly 
separated, the correct $K$ is rarely selected and ARI-based 
clustering performance is generally poor, so we focus our discussion 
on strongly separated cluster scenarios. Specifically, when the 
network is large or dense, the silhouette score is able to recover 
the correct $K$ for balanced cluster sizes, resulting in nearly 
perfect clustering accuracy accordingly. However, it fails to select 
the correct $K$ under the scenario of imbalanced cluster sizes. 
While clustering quality is poor in small, sparse networks, 
performance improves as networks grow larger and become denser, with 
the ARI median reaching as high as $0.9$ when $n = 600$ and 
$p_{\text{win}} = 0.6$.

\section{Example of Non-Convex Shapes}
\label{app:convex}

We present simple simulation to illustrate how the silhouette score 
behaves when applied to data with non-convex cluster shapes. The data 
consist of three concentric rings in a two-dimensional space (sample 
size $n = 600$), with each ring containing $200$ points. To construct 
the corresponding network adjacency, pairwise Euclidean distances are 
computed and rescaled to $[0,1]$ by subtracting the minimum distance 
and dividing by the overall range. The network adjacency matrix is 
then defined as $1$ minus the rescaled distance. 

\begin{figure}[H]
  \centering
  \begin{subfigure}[b]{0.45\textwidth}
    \centering
    \includegraphics[width=\textwidth]{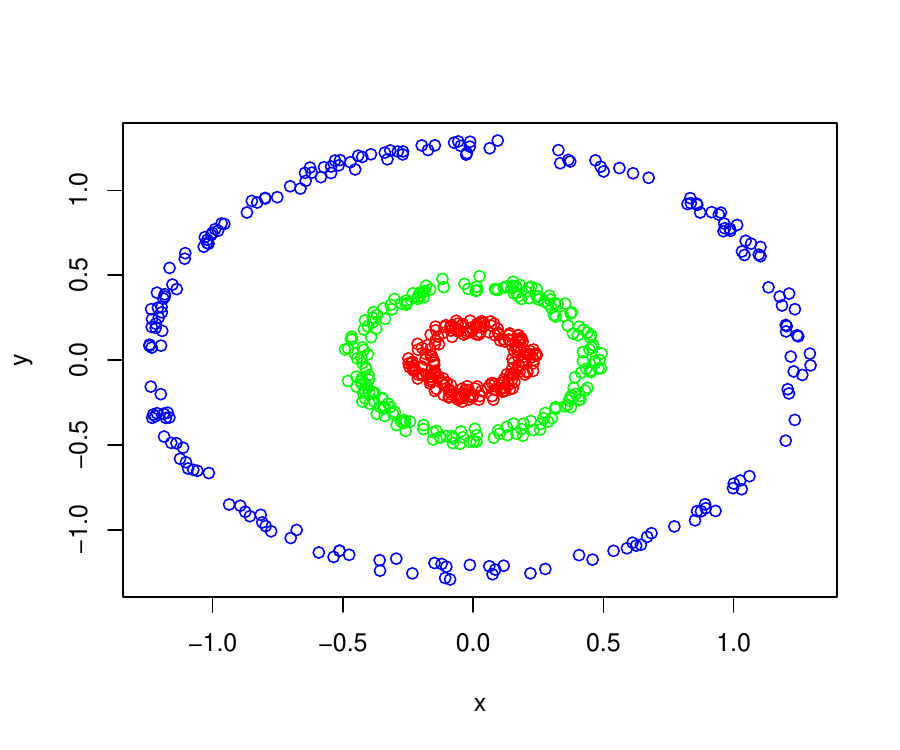}
  \end{subfigure}
  \hfill
  \begin{subfigure}[b]{0.45\textwidth}
    \centering
    \includegraphics[width=\textwidth]{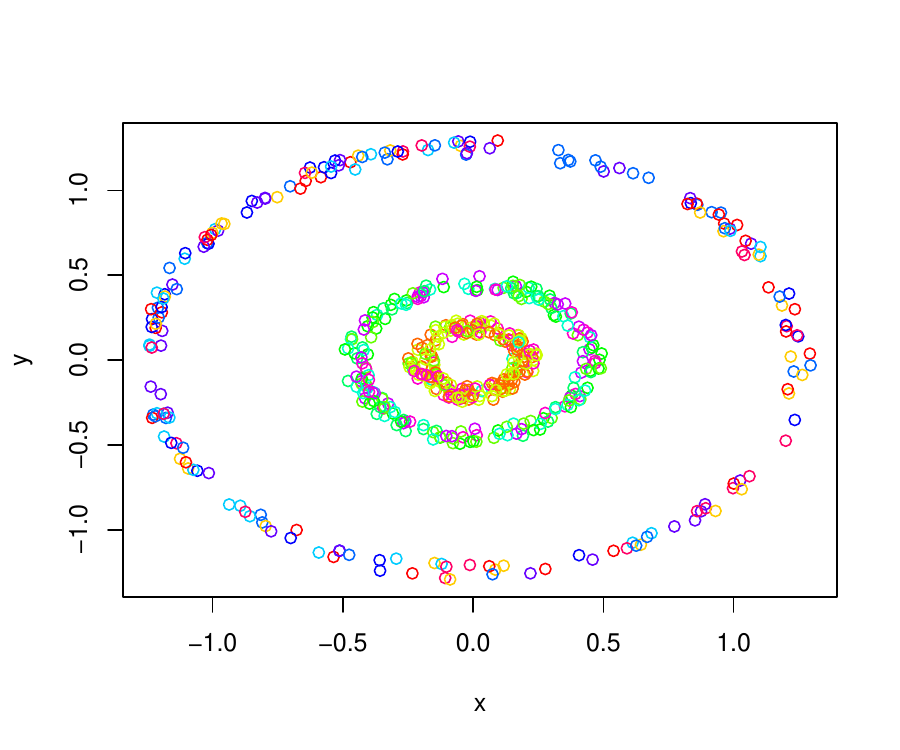}
      \end{subfigure}
  \caption{Simulated ring-shaped data with $K_\text{true} = 3$ (left) 
  and the clustering result selected by the silhouette score with $K 
  = 15$ (right), colored by cluster assignments.}
  \label{fig:ring}
\end{figure}

As shown in Figure~\ref{fig:ring}, the resulting clustering 
demonstrates that the silhouette score tends to subdivide each ring 
into multiple thinner rings. This occurs because, for non-convex 
structures such as concentric rings, points within the same true 
cluster may be far apart in Euclidean distance. Since the silhouette 
score rewards small within-cluster distances, it is inflated when 
each ring is partitioned into thinner rings or arc-shaped segments. 
This behavior aligns with our discussion in Section~\ref{sec:limit}, 
which highlights the limitations of the silhouette score when 
clusters deviate from convex geometry.

\end{document}